\documentclass[aps,twocolumn,superscriptaddress]{revtex4-1}

\usepackage{amsmath,amsfonts}
\usepackage{graphicx} 
\usepackage{epsfig}

\usepackage[printonlyused]{acronym}
\usepackage{multirow}
\usepackage{color}

\renewcommand{\vec}[1]{\underline{#1}}
\newcommand{\mat}[1]{\mathbf{#1}}

\newcommand{\Bracket}[3][ ]{\langle #2 | #1  #3\rangle}
\newcommand{\transpose}{\mathrm{t}}
\newcommand{\diag}{\mathrm{diag}}
\newcommand{\complexi}{\mathrm{i}}

\newcommand{\operator}[1]{\hat{#1}}
\newcommand{\Ket}[1]{|#1\rangle}

\newcommand{\ignore}[1]{}

\newcommand{\Trace}{\mathrm{Tr}}
\newcommand{\functional}[1]{\mathcal{#1}}

\def\noteC#1{\textbf{\color{green}}} 

\begin{document}

\title{Cumulant expansion for fast estimate of non-Condon effects in 
vibronic transition profiles}

\author{Joonsuk Huh}
\email{joonsukhuh@skku.edu}
\affiliation{Department of Chemistry, Sungkyunkwan University, Suwon 440-746, Korea}
\affiliation{Clemens-Sch\"opf Institute, TU Darmstadt, Petersenstr. 22, 64287 Darmstadt, Germany}
\affiliation{Frankfurt Institute for Advanced Studies, Goethe University, Ruth-Moufang-Str. 1, 60438 Frankfurt am Main,
Germany}
\author{Robert Berger}
\email{robert.berger@uni-marburg.de}
\altaffiliation[Present address: ]{Fachbereich Chemie, Philipps-Universit{\"a}t Marburg, Hans-Meerwein-Stra{\ss}e 4, 35032 Marburg,
Germany}
\affiliation{Clemens-Sch\"opf Institute, TU Darmstadt, Petersenstr. 22, 64287 Darmstadt, Germany}
\affiliation{Frankfurt Institute for Advanced Studies, Goethe University, Ruth-Moufang-Str. 1, 60438 Frankfurt am Main,
Germany}

\date{\today\ }

\begin{abstract}
When existing, cumulants can provide valuable information about a given
distribution and can in principle be used to either fully reconstruct or 
approximate the parent distribution function. A previously reported 
cumulant expansion approach for
Franck-Condon profiles [Faraday Discuss., 150, 363 (2011)] is extended to
describe also the profiles of vibronic transitions that are weakly allowed
or forbidden in the Franck-Condon approximation (non-Condon profiles). In
the harmonic approximation the cumulants of the vibronic profile
can be evaluated analytically and numerically with a coherent state-based
generating function that accounts for the Duschinsky effect. As
illustration, the one-photon
$1~^{1}\mathrm{A_{g}}\rightarrow1~^{1}\mathrm{B_{2u}}$ UV absorption
profile of benzene in the electric dipole and (linear) Herzberg-Teller
approximation is presented herein for zero Kelvin and finite temperatures.
\end{abstract}

\maketitle

\section{Introduction}
Vibrationally resolved electronic spectra (\emph{e.g.} one-photon
absorption and emission spectra) are within the Born-Oppenheimer framework
usually interpreted in terms of Franck-Condon (FC) factors (FCFs)
\cite{franck:1925,condon:1928}.  Accordingly, one can try to obtain the shape
of the spectral profile for a FC-allowed transition from computed FCFs in
frequency domain. However, the evaluation of FCFs for large molecular
systems is challenging even within the harmonic approximation if one has to
take Duschinsky mode mixing (rotation)~\cite{duschinsky:1937} into account. 
This is because multi-variate Hermite polynomials have then to be evaluated for
each FC integral, rather than only uni-variate Hermite polynomials as is
the case for the comparatively simple parallel harmonic oscillator model.  
The computational task becomes more difficult as the molecular size and
temperature increases because the number of FC integrals grows vastly. 
The evaluation of the multi-dimensional FC integral in the harmonic
approximation is, at least for some cases with Duschinsky mode mixing,
classified as a \#P-hard problem in computational complexity
theory~\cite{rahimi2015} and thus it recently became a topic in quantum
computation. A quantum optical simulation (quantum computation) has been
proposed theoretically for instance for the FC profile calculation and has
been performed for the photoelectron spectrum of SO$_{2}$ by a trapped-ion
device~\cite{huh2015,huh:VBS,Shen2017}.  

To describe FC-forbidden or weakly allowed transitions, one has to go
beyond the Condon approximation and employ for instance a Herzberg-Teller
(HT) expansion~\cite{herzberg:1933} of the electronic transition moment with
respect to the normal coordinates. As a result, the calculation of the
vibronic spectrum for a non-Condon process is even more difficult than for
a FC-allowed transition because one has to evaluate matrix elements of
the non-Condon operators which require for each HT integral in general the
calculation of combinations of several FC integrals.

The number of FC integrals and matrix elements of non-Condon operators to
be evaluated in a sum-over-states approach can, in principle, be significantly reduced with the help of integral configuration selection 
strategies~\cite{jankowiak:2007,santoro:2008}. However, this
time-independent (TI) calculation of the spectral profile in the frequency
domain is still considerably more expensive than an alternative
time-dependent (TD) approach that exploits time-correlation functions
(TCFs) (see \emph{e.g.} Ref.~\cite{tannor:1982,Borrelli2012,baiardi:2013}), but offers the ability to
directly assign individual peaks in the spectrum. As we have outlined
earlier~\cite{Huh2012}, a unified
coherent state-based generating function (CSGF)
approach~\cite{cjdoktorov:1979a} can be used both for rigorous integral
prescreening strategies and TCF calculations which combine the strengths of
both approaches and complement each other favorably. Even in the less
demanding TD approach, however, one usually invests significant
computational time for often unnecessary spectral details. 

Cumulants (or moments) of a distribution (see \emph{e.g.}
Refs.~\cite{lfcederbaum:1977,heller:1978a,cjdoktorov:1979a,islampour:1989,mukamel:1995,wadi:1999,schatz:2002,liang:2003,Lax1952,Kubo1955,tatchen:2008})
can deliver highly useful information. From this one can either attempt to
reconstruct the spectral shape or try to estimate the relevant spectral
profile, which can be exploited in subsequent TI and TD
approaches~\cite{Huh2011a,Huh2012}. Cumulants of the vibronic spectrum can be
obtained from the CSGF directly without computing the total spectrum in
frequency domain. This method was exploited already in
Ref.~\cite{huh:2011FD} for FC-allowed transitions, and we report herein an
extension of this method to incorporate non-Condon transitions. To illustrate the performance of the approach, we present the profile of the $^1\mathrm{A_{1g}}\rightarrow
~^1\mathrm{B_{2u}}$ transition of benzene, which frequently served as a prototypical example for multiple authors (see \emph{e.g.}~\cite{berger:1997a,he:2001,coriani:2010} and references therein). The transition is in the electric dipole
approximation Franck-Condon forbidden and it is studied herein at various temperatures within the
linear HT and harmonic approximation. Cumulant expansion is compared herein to the
TCF approach.     

\section{Method and computational detail}
\label{sec:method} The spectral profile ($\varrho(\hbar\omega;T)$) can be
expressed via the Fourier transform (FT) of the TCF ($\chi(t;T)$) that
depends on the time $t$ and temperature $T$, namely
\begin{align}
\varrho(\hbar\omega;T)=\hbar^{-1}\int_{-\infty}^{\infty}\mathrm{d}t 
~\chi(t;T)\mathrm{e}^{\complexi (\omega - \omega_{0}) t} , 
\end{align}
where $\hbar\omega$ is the transition energy and $\hbar\omega_{0}$ the
$0'-0$ transition energy.  The corresponding occupancy representation for
the TCF can be obtained from Fermi's Golden Rule, \emph{i.e.} 
\begin{align}
\chi(t;T)
=\tfrac{\sum_{\vec{v},\vec{v}'=\vec{0}}^{\infty} 
\langle\vec{v}'\vert\operator{\vec{\mu}}^{\dagger}(\vec{Q})\vert\vec{v}\rangle\langle\vec{v}'\vert\operator{\vec{\mu}}^{*}(\vec{Q})\vert\vec{v}\rangle^{*} \mathrm{e}^{-\complexi E_{\vec{\epsilon}',\vec{\epsilon}}t/\hbar}
\mathrm{e}^{-\vec{v}\cdot\vec{\epsilon}/(k_{\mathrm{B}}T)}}
{\prod_{k}(1-\mathrm{e}^{-\epsilon_{k}/(k_{\mathrm{B}}T)})^{-1}} \, ,
\label{eq:occxi}
\end{align}
where we have assumed an electric dipole transition with the electronic
transition dipole moment ($\vec{\operator{\mu}}(\vec{Q})$), which is a function
of normal coordinates of the initial electronic state, and the harmonic approximation.  The $N$-dimensional
harmonic oscillator eigenstates of the initial and final electronic state
are denoted by $\Ket{\vec{v}}=\Ket{v_{1},\ldots,v_{N}}$ and $\Ket{\vec{v}'}=\Ket{v_{1}',\ldots,v_{N}'}$ with the corresponding
harmonic energy vectors $\vec{\epsilon}=(\epsilon_{1},\ldots,\epsilon_{N})$ and $\vec{\epsilon'}=(\epsilon_{1}',\ldots,\epsilon_{N}')$,
respectively. $\operator{H}$ and $\operator{H}'$ are the $N$-dimensional harmonic oscillator Hamiltonians belonging to the initial and final
electronic states, respectively.
$k_{\mathrm{B}}$ is the Boltzmann constant.
$E_{\vec{\epsilon}',\vec{\epsilon}}$ is the vibronic transition energy with
respect to the $0'-0$ transition energy. 
The spatial representation of the TCF in closed form can be
found by evaluating the following quantum mechanical traces
\begin{align}
\chi(t;T)
=\tfrac{\Trace\Big(\operator{\vec{\mu}}(\vec{Q})^{\dagger}
 \exp(-\mathrm{i}\operator{H}'t/\hbar)
 \operator{\vec{\mu}}(\vec{Q})
\exp(\mathrm{i}\operator{H}t/\hbar)
\exp(-\operator{H}/(k_{\mathrm{B}}T))\Big)}
{\Trace (\mathrm{e}^{-\operator{H}/(k_{\mathrm{B}}T)})} \, .
\label{eq:closedform}
\end{align} 
The traces can be evaluated with any complete basis.  In our work $N$-dimensional coherent
states were used (see \emph{e.g.}
Refs.~\cite{jankowiak:2007,Huh2012}) with the
Duschinsky relation between initial and final state normal coordinates
($\vec{Q}'=\mat{S}\vec{Q}+\vec{d}$ where $\mat{S}$ and $\vec{d}$ are the
Duschinsky rotation matrix and displacement vector, respectively and
$\vec{Q}'$ are the normal coordinates of the final state) and the linear HT expansion
of the electronic transition dipole moment
($\operator{\vec{\mu}}\simeq\vec{\mu}(\vec{0})+\sum_{k}\vec{\mu}_{k}'\operator{Q}_{k}$
where $\vec{\mu}_{k}'$ is the first derivative of $\operator{\vec{\mu}}$
with respect to $\operator{Q}_{k}$.). 

The TCF is related to a probability
density function (PDF). If all cumulants or moments of a PDF are defined and 
available, the PDF can be reconstructed as follows~\cite{huh:2011FD}   
\begin{align}
\chi(t,T)=\varrho_{\mathrm{tot}} 
\exp\left(\sum_{k=1}^{\infty}\frac{\langle E_{\vec{\epsilon}',\vec{\epsilon}}^{k} \rangle^{\mathrm{c}}(T)}{k!}(\complexi t/\hbar)^{k}\right),
\end{align}
where $\langle E_{\vec{\epsilon}',\vec{\epsilon}}^{k}
\rangle^{\mathrm{c}}(T)$ is the $k$-th order cumulant at temperature $T$. 
The cumulants of the spectral density function are normalised to the total integrated profile $\varrho_{\mathrm{tot}}=\vert\vec{\mu}(\vec{0})\vert^{2}+\sum_{k}^{N}\frac{\hbar^{2}}{2\epsilon_{k}}\vert\vec{\mu}_{k}'\vert^{2}\coth (\frac{\epsilon_{k}}{2k_{\mathrm{B}}T})$~\cite{Huh2012}. 
Moments (cumulants and moments are inter-convertible~\cite{berberan-santos:2006}) 
can be obtained by partial derivatives of $\chi$ with respect to time, 
\begin{align} 
\langle E_{\vec{\epsilon}',\vec{\epsilon}}^{k} \rangle(T)
&=\Big(-\frac{\hbar}{\complexi}\Big)^{k}\dfrac{\partial^{k}}{\partial t^{k}}\chi(t;T) \Big\vert_{t=0}.  
\label{eq:moments}
\end{align}  

The cumulants can be obtained from the moments via the following transformation~\cite{berberan-santos:2006},
\begin{align}
\langle E_{\vec{\epsilon}',\vec{\epsilon}}^{n+1} \rangle^{\mathrm{c}} =
\langle E_{\vec{\epsilon}',\vec{\epsilon}}^{n+1} \rangle -
\sum_{k=0}^{n-1}\begin{pmatrix}n\\k\end{pmatrix}\langle
E_{\vec{\epsilon}',\vec{\epsilon}}^{n-k} \rangle\langle
E_{\vec{\epsilon}',\vec{\epsilon}}^{k+1} \rangle^{\mathrm{c}} . 
\end{align}

Thus, cumulants can be evaluated analytically or numerically by evaluating
partial derivatives of $\chi$ in Eq.~\eqref{eq:closedform} with respect to
the time variable at $t=0$. Analytic evaluation of the cumulants to
arbitrary order within the linear HT approximation can be performed along
the lines of the development in Refs.~\cite{huh:2011FD,Huh2011a,Huh2012}
for the cumulants of FC profiles to arbitrary order. For numerical evaluation 
of low-order cumulants
one needs to compute $\chi$ at the first few time steps. To obtain the
corresponding moments numerically, $\mathrm{Re}(\chi(t,T))$ and
$\mathrm{Im}(\chi(t,T))$ as computed at these time steps are used to determine low-order even and odd
moments, respectively, because
$\mathrm{e}^{-\complexi
E_{\vec{\epsilon}',\vec{\epsilon}}t/\hbar}=\cos(E_{\vec{\epsilon}',\vec{\epsilon}}t/\hbar)-\complexi
\sin(E_{\vec{\epsilon}',\vec{\epsilon}}t/\hbar)$ in Eq.~\eqref{eq:occxi}
(see \emph{e.g.} Ref.~\cite{schatz:2002}).

The closed form of $\chi(t,T)$ within the linear HT approximation can be found 
in Refs.~\cite{Huh2011a,Huh2012,Borrelli2012,baiardi:2013}. In the present work, flexibility
is used in the GF to
obtain detailed information concerning individual contributions of different modes. This is achieved by assigning different time and temperature
variables to each vibrational degree of freedom. The corresponding GF in
an occupancy representation reads as follows
\begin{align} 
G^{K}(\mat{Z};\mat{\tilde{\Gamma}})^{(\operator{f},\operator{g})} 
=&\mathcal{N}|\Bracket{\vec{0}'}{\vec{0}}|^{-2}\sum_{\vec{v},\vec{v}'=\vec{0}}^{\vec{\infty}}
\langle\vec{v}'\vert\operator{f}\vert\vec{v}\rangle
\langle\vec{v}'\vert\operator{g}\vert\vec{v}\rangle^{*}  \nonumber \\
&\prod_{k=1}^{N} [ z_{k}^{2v_{k}}(z_{k}')^{2v_{k}'} ]
\mathrm{e}^{-(\vec{v}^\transpose\mat{B}\vec{\epsilon}
              +\vec{v}'^\transpose\mat{B}'\vec{\epsilon}')} \, ,
\label{eq:occfg}
\end{align}  
where the general operators $\operator{f}$ and $\operator{g}$, which can be
products of momentum and position operators, are given instead of 
$\operator{\vec{\mu}}(\operator{\vec{Q}})$; and different temperatures can be given to the initial and final
vibrational degrees of freedom via
\begin{align}
\mat{B}=\mathrm{diag}(\beta_{1},\ldots,\beta_{N}),~\mat{B}'=\mathrm{diag}(\beta_{1}',\ldots,\beta_{N}'), 
\end{align}
where $\beta_{k}=1/(k_{\mathrm{B}}T_{k})$ (with Boltzmann constant $k_{\mathrm{B}}$ and temperature $T_{k}$).
The parameter matrices are defined as follows
\begin{align}
\mat{Z}&=\begin{pmatrix} \mat{z} & \mat{0} \\ \mat{0} & \mat{z}'
\end{pmatrix} \, ,
\end{align}
with the time variables being assigned to
the matrices $\mat{z}$ = $\diag(\mathrm{e}^{\complexi
\epsilon_{1}t_{1}/(2\hbar)},\ldots,\mathrm{e}^{\complexi
\epsilon_{N}t_{N}/(2\hbar)})$ and $\mat{z}'$ =
$\diag(\mathrm{e}^{-\complexi
\epsilon_{1}'t_{1}'/(2\hbar)},\ldots,\mathrm{e}^{-\complexi
\epsilon_{N}'t_{N}'/(2\hbar)})$ for initial and final vibrational modes,
respectively, 
and
\begin{align}
\mat{\widetilde{\Gamma}}&=\begin{pmatrix} \mat{\Gamma} & \mat{0} \\ \mat{0} & \mat{\Gamma}'
\end{pmatrix} \,, 
\end{align}
with 
$\mat{\Gamma}=\diag(\mathrm{e}^{-\beta_{1}\epsilon_{1}/2},\ldots,\mathrm{e}^{-\beta_{N}\epsilon_{N}/2} ), 
\mat{\Gamma}'=\diag(\mathrm{e}^{-\beta_{1}'\epsilon_{1}'/2},
\ldots,\mathrm{e}^{-\beta_{N}'\epsilon_{N}'/2} ) $. 

 $\mathcal{N}$ is the corresponding normalizing factor related to the partition function of the Boltzmann distribution of harmonic oscillators, i.e.       
\begin{align}
\mathcal{N}=\prod_{k}^{N}(1-\mathrm{e}^{-\beta_{k}\epsilon_{k}})(1-\mathrm{e}^{-\beta_{k}'\epsilon_{k}'}) .
\end{align}

The Duschinsky relation is considered with the Doktorov matrices and vectors \cite{doktorov:1977}
\begin{align}
\mat{W}&=\begin{pmatrix}\mat{I}-2\mat{Q} & -2\mat{R} \\
-2\mat{R}^{\transpose} & \mat{I}-2\mat{P}\end{pmatrix} \, , \quad
\vec{r}=\sqrt{2}\begin{pmatrix}-\mat{R}\vec{\delta} \\
(\mat{I}-\mat{P})\vec{\delta}\end{pmatrix} \, ,
\label{eq:Wmatrix}
\end{align}

\begin{align} 
\mat{Q}&=(\mat{I}+\mat{J}^\transpose\mat{J})^{-1}\, ,\quad
\mat{P}=\mat{J}\mat{Q}\mat{J}^\transpose \, , \nonumber \\
\mat{R}&=\mat{Q}\mat{J}^\transpose\, ,\quad \mat{J}=\mat{\Omega}'\, \mat{S} \,
\mat{\Omega}^{-1}\, , \quad\vec{\delta}=\mat{\Omega}'\vec{d}/\sqrt{\hbar} \, ,
\end{align}
as well as
\begin{align}
\mat{\Omega}&=\diag(\vec{\epsilon})^{1/2}/\sqrt{\hbar},  \nonumber \\
\mat{\Omega}'&=\diag(\vec{\epsilon}')^{1/2}/\sqrt{\hbar}
\, .
\end{align}
The temperature dependent parameters are defined as follow
\begin{align}
\mat{W}_{T}=&\mat{\widetilde{\Gamma}}\mat{W}\mat{\widetilde{\Gamma}}, \quad
\vec{r}_{T}=\mat{\widetilde{\Gamma}} \vec{r} .
\end{align}

Finally, the Franck-Condon Herzberg-Teller (FCHT) TCF reads as 
\begin{align} 
\frac{\chi_{\mathrm{FCHT}}(\mat{Z};\mat{\tilde{\Gamma}})}{|\Bracket{\vec{0}'}{\vec{0}}|^{2}}
=
&\vert\vec{\mu}_{0}\vert^{2}G^{K}(\mat{Z};\mat{\tilde{\Gamma}}) \nonumber \\
&+2\sum_{i}\vec{\mu}_{0}\cdot\vec{\mu}_{i}'G^{K}(\mat{Z};\mat{\tilde{\Gamma}})^{
(\operator{Q}_{i},\operator{1})} \nonumber \\
&+\sum_{i,j}\vec{\mu}_{i}'\cdot\vec{\mu}_{j}'G^{K}(\mat{Z};\mat{\tilde{\Gamma}})^{
(\operator{Q}_{i},\operator{Q}_{j})}
\label{eq:FCHTgenfunc} \, , 
\end{align}
with the FC generating function,  
\begin{align}
&G^{K}(\mat{Z};\mat{\widetilde{\Gamma}})= G^{K}(\mat{Z};\mat{\widetilde{\Gamma}})^{(\hat{1},\hat{1})} \nonumber \\  &\mathcal{N}\mathrm{det}(\mat{I}+\mat{Z}\mat{W}_{T}\mat{Z})^{-1/2}\mathrm{det}(\mat{I}-\mat{Z}\mat{W}_{T}\mat{Z})^{-1/2} \nonumber \\ 
&\exp((\vec{r}_{T})^\transpose\mat{Z}(\mat{I}+\mat{Z}\mat{W}_{T}\mat{Z})^{-1}\mat{Z}\vec{r}_{T}) ,
\end{align}
the mixed FC/HT generating function,
\begin{align}
&G^{K}(\mat{Z};\mat{\widetilde{\Gamma}})^{(\hat{Q}_{i},\hat{1})}=  \nonumber \\ 
&=\sqrt{\tfrac{\hbar}{2(\epsilon_{i}/h)}}G^{K}(\mat{Z};\mat{\widetilde{\Gamma}}) \nonumber \\ 
&[\vec{r}+(\mat{I}-\mat{W})\mat{\widetilde{\Gamma}}\mat{Z}(\mat{I}+\mat{Z}\mat{W}_{T}\mat{Z})^{-1}\mat{Z}\vec{r}_{T}]_{i} \label{eq:fchtfgeneratingfunction} \, ,
\end{align}
and the HT generating function,
\begin{align}
&G^{K}(\mat{Z};\mat{\widetilde{\Gamma}})^{(\hat{Q}_{i},\hat{Q}_{j})}=\nonumber \\ 
&=\tfrac{\hbar}{2}\sqrt{\tfrac{1}{(\epsilon_{i}/h)(\epsilon_{j}/h)}}G^{K}(\mat{Z};\mat{\widetilde{\Gamma}}) \nonumber \\  
&\Big[[\vec{r}+(\mat{I}-\mat{W})\mat{\widetilde{\Gamma}}\mat{Z}(\mat{I}+\mat{Z}\mat{W}_{T}\mat{Z})^{-1}\mat{Z}\vec{r}_{T}]_{i} \nonumber \\  
&\times[\vec{r}+(\mat{I}-\mat{W})\mat{\widetilde{\Gamma}}\mat{Z}(\mat{I}+\mat{Z}\mat{W}_{T}\mat{Z})^{-1}\mat{Z}\vec{r}_{T}]_{j} \nonumber \\  
&+\tfrac{1}{2}[(\mat{I-W})\mat{\widetilde{\Gamma}}\mat{Z}(\mat{I}+\mat{Z}\mat{W}_{T}\mat{Z})^{-1}
\mat{Z}\mat{\widetilde{\Gamma}}(\mat{I-W})]_{ij} \nonumber \\ 
&+\tfrac{1}{2}[(\mat{I-W})\mat{\widetilde{\Gamma}}\mat{Z}(\mat{I}-\mat{Z}\mat{W}_{T}\mat{Z})^{-1}
\mat{Z}\mat{\widetilde{\Gamma}}(\mat{I-W})]_{ji}\Big] \, .
\end{align}

The electronic $1~^{1}\mathrm{A_{g}}\rightarrow1~^{1}\mathrm{B_{2u}}$
transition of benzene is FC-forbidden in the electric dipole approximation
($\vec{\mu}(\vec{0})=\vec{0}$) such that only the HT terms contribute to
the spectral function. The corresponding TCF for FCHT weighted density of states (FCHTW) is given as follows,
here with same time ($t$) for all vibrational degrees of
freedom and same temperature ($T_{k}=T$ and $T_{k}'=\infty$) for the initial and final modes, respectively, 
\begin{align}
\chi_{\mathrm{FCHTW}}(t;T)=|\Bracket{\vec{0}'}{\vec{0}}|^{2}
\sum_{i,j}\vec{\mu}_{i}'\cdot\vec{\mu}_{j}'G(t;T)^{(\operator{Q}_{i},\operator{Q}_{j})}.
\end{align}
Accordingly, the spectrum is obtained by the Fourier transformation, 
\begin{align}
\varrho_{\mathrm{FCHTW}}(\hbar\omega;T)=\hbar^{-1}\int_{-\infty}^{\infty}\mathrm{d}t 
~\chi_{\mathrm{FCHTW}}(t;T)\mathrm{e}^{\complexi (\omega - \omega_{0}) t} , 
\end{align}
which can show the detailed vibronic structure. 

The FCHTW profile is now approximated with a finite number of cumulants via the Edgeworth expansion with the order $n$~\cite{blinnikov:1998}. Whereas for $n=2$
a Gaussian distribution function is used, the Edgeworth expansion for order $n \geq 3$ is employed as~\cite{huh:2011FD}
\begin{align}
&\varrho_{\rm{FCHTW}}^{(\mathrm{c})}(\tilde{\nu};T;n\geq 3)\nonumber \\
&=
\frac{\rho_{\mathrm{tot}}}{\sqrt{2\pi(hc_{0})^{-2}\langle E_{\vec{\epsilon}',\vec{\epsilon}}^{2}
\rangle^{\mathrm{c}}(T)}} 
\exp\Big(-\frac{(\tilde{\nu}-\tilde{\nu}_{0})^2}{2(hc_{0})^{-2}\langle
E_{\vec{\epsilon}',\vec{\epsilon}}^{2} \rangle^{\mathrm{c}}(T)}\Big)
\nonumber \\
&\Big[ 1+\sum_{s=1}^{n}\Big(\sqrt{(hc_{0})^{-2}\langle
E_{\vec{\epsilon}',\vec{\epsilon}}^{2}
\rangle^{\mathrm{c}}(T)}\Big)^{s}\nonumber \\
&\times\sum_{\{\vec{k}\}}\functional{H}_{s+2r}(\tilde{\nu}-\tilde{\nu}_{0})
\prod_{m=1}^{s}\frac{1}{k_{m}!}\Big(\frac{S_{m+2}(T)}{(m+2)!}\Big)^{k_{m}}\Big],
\label{eq:edgeworthex}
\end{align}
where $\{\vec{k}\}$ is a set of non-negative integer vectors, which are constrained to    
$s=\sum_{m=1}^{s}m k_{m}$ and $r=\sum_{m=1}^{s}k_{m}$. 
$\functional{H}_{s+2r}$ is a uni-variate Hermite polynomial of order $s+2r$ and $\mathcal{S}_{m+2}$ is defined as follows,
\begin{align}
\mathcal{S}_{m+2}(T)=
\frac{(hc_{0})^{-(m+2)}\langle E_{\vec{\epsilon}',\vec{\epsilon}}^{m+2}
\rangle^{\mathrm{c}}(T)}{((hc_{0})^{-2}\langle E_{\vec{\epsilon}',\vec{\epsilon}}^{2}
\rangle^{\mathrm{c}}(T))^{m+1}} .
\end{align}
The Edgeworth expansion with a finite number of cumulants and in the infinite series are related as 
\begin{align}
\lim_{n\rightarrow \infty}\varrho_{\rm{FCHTW}}^{(\mathrm{c})}(\tilde{\nu};T;n)
= \varrho_{\rm{FCHTW}}(\tilde{\nu};T), 
\end{align}
and we use the relation, $\varrho_{\rm{FCHTW}}(\tilde{\nu};T)=hc_{0}\varrho_{\rm{FCHTW}}(\hbar\omega;T)$, 
for the wavenumber ($\tilde{\nu}$) domain profile.  

\label{sec:TICEresult}
\begin{figure}
\begin{center}
\includegraphics[width=0.45\textwidth]{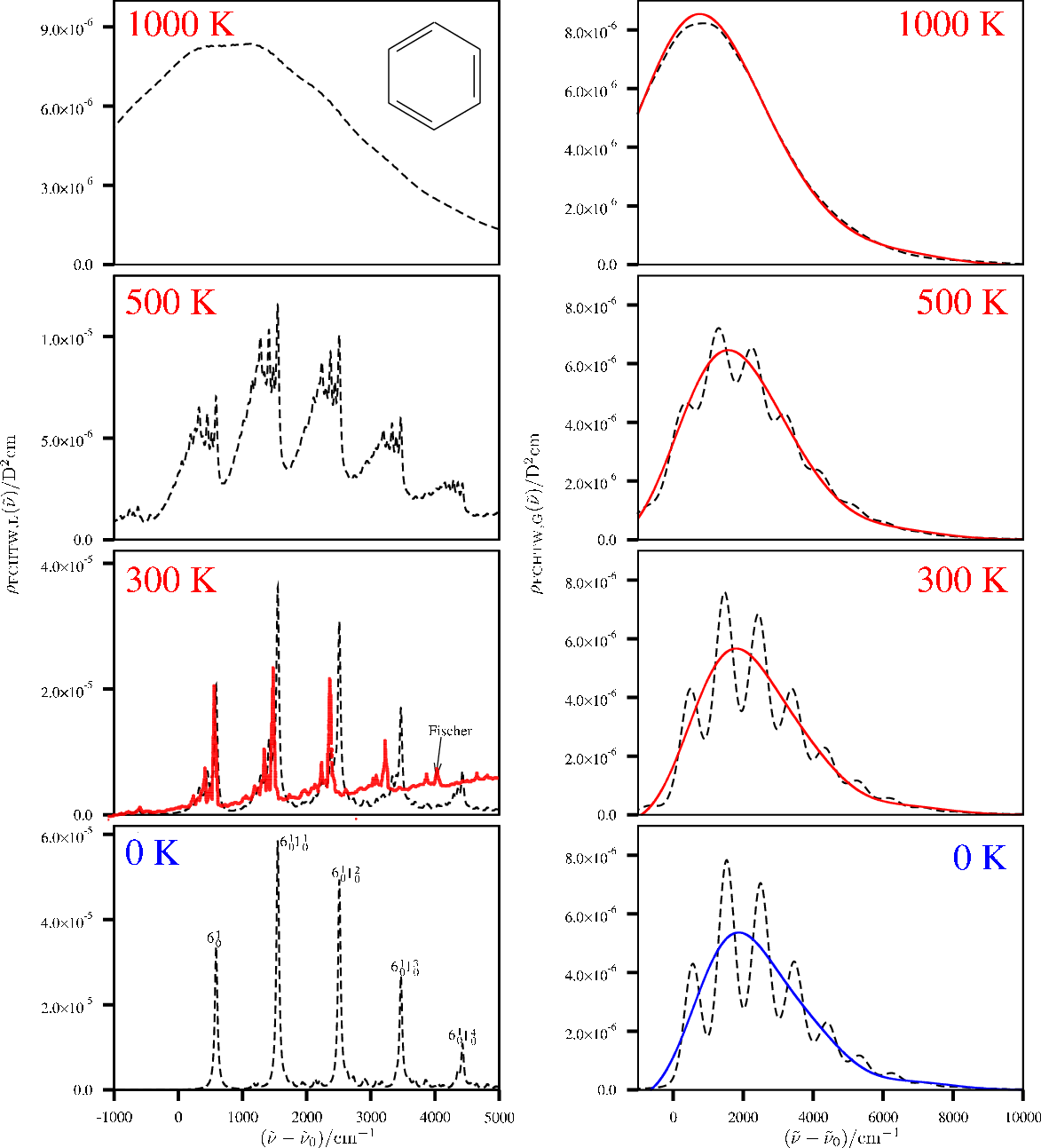} 
\end{center}
\caption{Left part of the figure: The dashed lines are drawn for the
TCF-FFT approach with a Lorentzian line shape function with FWHM of
50~cm$^{-1}$. A time increment $\Delta t$ of 0.51~fs and a grid with
2$^{16}$ grid points are used for the corresponding FFT calculations. The
experimental UV absorption spectrum as reported by Fischer in
Ref.~\cite{fischer} is additionally shown in red, which has been shifted to
match approximately the position of the major peak in the region below the
0-0 transition wavenumber and rescaled to have similar peak height as the
one computed for the $6_{0}^{1}$ transition. Right part of the figure: The
dashed lines are drawn for the TCF-FFT approach with a Gaussian line shape
function of with FWHM of 500~cm$^{-1}$.  A time increment of 0.10~fs and a
grid with 2$^{15}$ grid points are used for the corresponding FFT
calculations.  Solid lines are drawn for the curve obtained by Edgeworth
expansion using up to 4-th order cumulants and, for 1000~K by Edgeworth
expansion using up to 8-th order cumulants.  
}\label{fig:benzenespectrum}
\end{figure}
The vibronic profiles for benzene's
$1~^{1}\mathrm{A_{g}}\rightarrow1~^{1}\mathrm{B_{2u}}$ transition at zero
Kelvin and finite temperatures are calculated with the two methods, namely TCF
and time-independent cumulant expansion (CE). 
We use herein the term time-independent CE, which was employed in 
Ref.~\cite{huh:2011FD}, to distinguish this CE from the conventional 
(time-dependent) CE (see \emph{e.g.} 
Refs.~\cite{islampour:1989,mukamel:1995,wadi:1999,schatz:2002,liang:2003}) 
which involves time integration for the cumulant calculation.
To compute the vibronic spectra via the TCF
method, the FFTW library~\cite{FFTW05} is used for fast Fourier Transform
(FFT). The approximate curves are generated for the CE with Edgeworth
expansion~\cite{blinnikov:1998,huh:2011FD} using the computed low-order 
cumulants. Some of the problems related to this type of expansion for the
description of FC profiles are discussed in Ref.~\cite{huh:2011FD}. The 
moments (Eq.~\ref{eq:moments}) are calculated both analytically 
and numerically for comparison, the latter by approximating partial
derivatives of $\chi$ in Eq.~\ref{eq:closedform} with respect to time (see
results section) via a central finite difference scheme with a truncation
error being of order $(\Delta t)^2$.
When generating the data points in time, we exploit
the time-reversal symmetry condition, i.e. $\chi(-t;T)=\chi(t;T)^{*}$.
Required input data from
electronic structure calculations for benzene, \emph{i.e.} molecular
equilibrium structures and corresponding harmonic force fields for each
electronic state ($^{1}\mathrm{A_{g}}$ and $^{1}\mathrm{B_{2u}}$) as well
as first derivatives of the electronic transition dipole moments are taken
from Ref.~\cite{berger:1997a} (CASSCF/DZV). These data have been 
compared to results obtained via analytical derivative techniques for 
electronic transition dipole moments within a time-dependent density 
functional theory framework in Ref.~\cite{coriani:2010}.
The vibronic structure methods employed in the present work are
implemented in a development version of our vibronic structure program
package
hotFCHT~\cite{berger:1997a,jankowiak:2007}. 

\section{Results and discussion}
The computed vibronic spectra are shown in Fig.~\ref{fig:benzenespectrum}. 
The left hand side in Fig.~\ref{fig:benzenespectrum} shows vibronic
profiles from TCF-FFT which are convoluted by a Lorentzian line shape
function with full width at half maximum (FWHM) of 50~cm$^{-1}$ at
temperatures elevating from 0~K to 1000~K. This FC-forbidden vibronic
transition is mediated by the non-totally symmetric vibrational modes in
the irreducible representation e$_{\mathrm{2g}}$ of the D$_\mathrm{2h}$ 
molecular symmetry group. The main feature of the
vibronic spectrum is from progressions in the totally symmetric C-C
stretching mode (963 cm$^{-1}$) building on the so-called false origin from
a single excitation of a non-totally symmetric (e$_{\mathrm{2g}}$) in-plane
bending mode (575 cm$^{-1}$) as indicated in the spectrum at zero Kelvin.
The calculated spectrum at 300 K is compared with the experimental data of
Fischer~\cite{fischer}. The two spectra agree fairly well in the low energy
region but the computed peaks at higher energies are slightly
shifted to larger wavenumbers due to the harmonic approximation. As
temperature increases the vibrational structure becomes very congested and
washed out. At 1000 K (only employed for testing the method) one can not see a
resolved vibrational structure any longer, only the corresponding envelope.

On the right hand side of Fig.~\ref{fig:benzenespectrum} the two methods
(TCF-FFT and CE-Edgeworth) are compared for increasing temperatures. The
spectra are convoluted in the TCF-FFT curves (dashed lines) with a Gaussian
line shape function of 500~cm$^{-1}$ for FWHM. The second moments
(4.51$\times$10$^{4}$ $\mathrm{cm}^{-2}~(hc_0)^{2}$) of the Gaussian line
shape function is added to the second moments of the vibronic spectrum to
take the line shape function into account (see
Ref.~\cite{huh:2011FD} for the rationalisation and details). The relatively
broad line shape function is used for the TCF-FFT curves to have
vibrationally relatively structureless spectra for comparison. At 0, 300
and 500 K, the TCF-FFT curves still show vibrational structure and the
CE-Edgeworth curves (solid lines) look like nonlinear regression curves of
the corresponding TCF-FFT spectra. When the vibrational structures are also
essentially smoothed out in the TCF-FFT curves at 1000 K, the two
approaches agree with each other extremely well. Up to the 4-th order
cumulants are used for 0, 300, 500 K and up to 8-th order cumulants are
computed for 1000 K. 

In table~\ref{table:moments} the moments computed numerically (via
numerical derivatives) and analytically are compared. At low orders and all
temperatures the two methods agree well and for higher orders still the
agreement is satisfactory. One of the advantages of the numerical method is
that one only needs to compute the TCF for the first few time steps and it 
can be improved by controlling the time increment and the number of data points. The
analytical method usually meets a combinatorial problem in high order
cumulant calculations due to the analytic derivatives of the inverse
matrix~\cite{huh:2011FD}.  The second advantage of the numerical 
method is that it is easy to include linear and
nonlinear non-Condon effects. The third advantage is that one can
incorporate general line shape functions which would not have well defined
cumulants (see the discussion on page 415 of Ref.~\cite{huh:2011FD}). Lastly, 
the computational cost of the numerical cumulant expansion method is that 
the number of data points to be evaluated is almost negligible comparing to 
the TCF-FFT approach, which is about three orders of magnitude more expensive.

\begin{table}[hbt]
\caption{Analytically and numerically computed cumulants.
4.51$\times$10$^{4}$ $\mathrm{cm}^{-2}~(hc_0)^{2}$ is added to the second
moments to take the Gaussian line shape function (FWHM = 500 cm$^{-1}$)
into account; see Ref.~\cite{huh:2011FD} for details. A time increment of
0.10~fs is used for computing the numerical derivatives.}
\label{table:moments}
\centering
\scalebox{0.65}{
    \begin{tabular}{cllllllll} 
\hline
\multirow{2}{*}{$n$} &\multicolumn{8}{c}{$\langle E_{\vec{\epsilon}',\vec{\epsilon}}^{n} \rangle/(\mathrm{cm}^{-1}~hc_0)^{n}$} \\
\cline{2-9}
       &        \multicolumn{2}{c}{$T=0~\mathrm{K}$}     & \multicolumn{2}{c}{$T=300~\mathrm{K}$} & 
                \multicolumn{2}{c}{$T=500~\mathrm{K}$}   & \multicolumn{2}{c}{$T=1000~\mathrm{K}$}\\
\cline{2-9}
& Analytical & Numerical & Analytical & Numerical & Analytical & Numerical & Analytical & Numerical\\
\hline
1   &  2.61$\times 10^{3}$ &2.61$\times 10^{3}$ &2.47$\times 10^{3}$ &2.47$\times 10^{3}$ &2.12$\times 10^{3}$ &2.12$\times 10^{3}$ &1.12$\times 10^{3}$ &1.12$\times 10^{3}$ \\
2   &  9.24$\times 10^{6}$ &9.21$\times 10^{6}$ &8.64$\times 10^{6}$ &8.62$\times 10^{6}$ &7.38$\times 10^{6}$ &7.36$\times 10^{6}$ &5.64$\times 10^{6}$ &5.63$\times 10^{6}$  \\
3   &  4.07$\times 10^{10}$&4.05$\times 10^{10}$&3.77$\times 10^{10}$&3.75$\times 10^{10}$&3.17$\times 10^{10}$&3.15$\times 10^{10}$&2.03$\times 10^{10}$&2.01$\times 10^{10}$ \\
4   &  2.14$\times 10^{14}$&2.12$\times 10^{14}$&1.97$\times 10^{14}$&1.95$\times 10^{14}$&1.66$\times 10^{14}$&1.64$\times 10^{14}$&1.29$\times 10^{14}$&1.27$\times 10^{14}$ \\
5   &     -         &     -     &  -   &     -         &     -     &  -        &                                                     8.00$\times 10^{17}$ &7.82$\times 10^{17}$\\
6   &     -         &     -     &  -   &     -         &     -     &  -        &                                                     6.55$\times 10^{21}$ &6.33$\times 10^{21}$\\
7   &     -         &     -     &  -   &     -         &     -     &  -        &                                                     5.84$\times 10^{25}$&5.54$\times 10^{25}$\\
8   &     -         &     -     &  -   &     -         &     -     &  -        &                                                     6.16$\times 10^{29}$&5.71$\times 10^{29}$\\
\hline
\end{tabular}
}
\end{table}

\begin{table}[hbt]
\caption{Mean excitation wavenumbers of the components of individual
vibrational $\mathrm{e_{2g}}$ symmetric modes of benzene as computed for
different tempertures.  The numbering used for the modes $\nu_6$, $\nu_7$,
$\nu_8$ and $\nu_9$, correspond to that used by Wilson for benzene and
translates to $\nu_{18}$, $\nu_{15}$, $\nu_{16}$ and $\nu_{17}$ in
Herzberg's nomenclature, respectively. The corresponding harmonic vibrational 
wavenumbers $\tilde{\omega}_\mathrm{e}'$ as computed in
Ref.~\cite{berger:1997a} for the electronically excited state
and as used in the present calculations are also given.}
\label{table:mean}
\scalebox{0.7}{
\centering
\begin{tabular}{llllll}
\hline
         &       & \multicolumn{4}{c}{$\epsilon_i'\langle\hat{v}_i'\rangle/(hc_0~\mathrm{cm}^{-1})$}\\
Mode $\nu_i$ & $\tilde{\omega}_\mathrm{e}'$/cm$^{-1}$&$T$ = 0~K &  $T$ = 300 K         &  $T$ = 500 K         &  $T$ = 1000 K        \\
\hline
$\nu_6$ & 575   &2.62$\times 10^{2}$ &  2.90$\times 10^{2}$ &  3.79$\times 10^{2}$ &  7.25$\times 10^{2}$ \\
$\nu_6$ & 575   &2.62$\times 10^{2}$ &  2.90$\times 10^{2}$ &  3.79$\times 10^{2}$ &  7.25$\times 10^{2}$ \\
$\nu_9$ & 1237  &1.60$\times 10^{1}$ &  1.78$\times 10^{1}$ &  4.64$\times 10^{1}$ &  2.51$\times 10^{2}$ \\
$\nu_9$ & 1237  &1.60$\times 10^{1}$ &  1.78$\times 10^{1}$ &  4.64$\times 10^{1}$ &  2.51$\times 10^{2}$ \\
$\nu_8$ & 1665  &2.54$\times 10^{1}$ &  2.38$\times 10^{1}$ &  3.09$\times 10^{1}$ &  1.65$\times 10^{2}$ \\
$\nu_8$ & 1665  &2.54$\times 10^{1}$ &  2.38$\times 10^{1}$ &  3.09$\times 10^{1}$ &  1.65$\times 10^{2}$ \\
$\nu_7$ & 3389  &8.14$\times 10^{1}$ &  7.49$\times 10^{1}$ &  6.08$\times 10^{1}$ &  6.37$\times 10^{1}$ \\
$\nu_7$ & 3389  &8.14$\times 10^{1}$ &  7.49$\times 10^{1}$ &  6.08$\times 10^{1}$ &  6.37$\times 10^{1}$ \\
\hline
\end{tabular}
}
\end{table}

Mean excitation wavenumbers ($\epsilon_{i}'\langle\hat{v}_{i}'\rangle/(hc_0)$ with 
$\hat{v}_{i}'$ being a number operator of $i$-th mode in the final electronic 
state) of individual modes are computed analytically for the HT active 
$\mathrm{e_{2g}}$ symmetric vibrational modes of the final state, and are 
given in table~\ref{table:mean}. The corresponding first derivative in 
Eq.~\eqref{eq:moments} can be performed numerically or analytically by 
assigning $\mat{z}=\diag(1,\ldots,1)$ and $\mat{z'}=\diag(1,\ldots,1,\mathrm{e}^{\mathrm{i}\epsilon_{i}'t/(2\hbar)},1,\ldots{1})$ to Eq.~\ref{eq:FCHTgenfunc}. The mean excitation
energy can serve as a parameter for the individual vibrational degrees of
freedom as an effective reorganisation energy or a Huang--Rhys factor (when
normalised by its harmonic energy), which can be characterised as a
function of structural deformation, frequency change, Duschinsky mode
coupling and temperature both in the Condon and non-Condon approximation.
One might naively expect a larger mean excitation energy as the temperature
increases but the mean values of high frequency modes (1665 and 3389
cm$^{-1}$) in some intermediate temperature rages are smaller than at zero
Kelvin. This can be rationalised as follows: Because the Duschinsky mode
mixing between low and high frequency modes is small in the
present case, the high frequency modes can not obtain thermal energy from the 
low frequency modes efficiently. Thus the high frequency modes are almost 
thermally inactive, whereas the total integrated
profile ($\varrho_{\mathrm{tot}}$) increases as temperature increases. In
the mean energy calculation of the high frequency modes at finite
temperatures the denominator (total intensity) increases because low
frequency modes accept thermal energy while the numerator (excitation of
high frequency modes) stays constant. Therefore the mean excitation
energies of high frequency modes are reduced at finite temperatures.
If Duschinsky rotation couples the low and high frequency
modes significantly, however, 
thermal energy can be transfered to the high
frequency modes via the low frequency modes in the initial state,
accordingly the mean excitation energies of high frequency modes can
increase as temperature increases.   

In closing the section, the moments or the cumulants of the vibronic
excitation energy can provide intuitively useful information concerning the
vibronic transition profile with almost no computation cost comparing to
the TCF-FFT method. Furthermore, the mean excitation energy of individual
mode opens a new interpretation for the vibronic transition with a single
quantity incorporating the mode mixing and the non-Condon effects as well
as the temperature, geometrical change and distortion effects.  

\section{Conclusion and outlook} 
\label{sec:TICEconclusion} 
We have
discussed a cumulant expansion method for describing non-Condon transitions
and applied it to the prototypical one-photon electric dipole
$1~^{1}\mathrm{A_{g}}\rightarrow1~^{1}\mathrm{B_{2u}}$ transition of
benzene, which is FC forbidden but HT allowed in the linear HT
approximation.  The method is particularly powerful when one does not
require all the details of the vibronic structures, but rather only
quantities such as peak maximum, mean and variance of the spectral shape.
This method is computationally much cheaper than the sum-over-states and
time-correlation function approach.  Moreover, the information (\emph{e.g.}
on the spectroscopically relevant energy window) from the cumulant expansion 
method can be used in the calculation within the other two methods. Herein we 
compared a numerical approach for the calculation of cumulants with the 
results from an analytical scheme. The results obtained numerically are still
fairly good. With this method, one can incorporate easily nonlinear
non-Condon terms and various line shape functions. In the
time-correlation function calculation the real part and imaginary part at
each time step provide automatically the even and odd moments,
respectively. In the first few time steps we already have the first few
moments available and the probability distribution function (information)
becomes complete as time progresses.  
The benzene example selected herein serves to illustrate the principle of 
the method for future routine applications for molecular systems with 
hundreds of atoms.  
 
\section*{Acknowledgements}
Financial support by the Beilstein-Institut, Frankfurt/Main, and computer time 
provided by the Center for Scientific Computing (CSC) Frankfurt are gratefully 
acknowledged. We are indebted to Jason Stuber for discussions. J.H. acknowledges 
supports by Basic Science Research Program through the National Research 
Foundation of Korea (NRF) funded by the Ministry of Education, Science and 
Technology (NRF-2015R1A6A3A04059773).
\bibliographystyle{apsrev4-1}
%

\end{document}